\documentstyle[12pt,epsfig]{article}

\catcode`@=11
\newcount\@tempcntc
\def\@citex[#1]#2{\if@filesw\immediate\write\@auxout{\string\citation{#2}}\fi
  \@tempcnta\z@\@tempcntb\m@ne\def\@citea{}\@cite{\@for\@citeb:=#2\do
    {\@ifundefined
       {b@\@citeb}{\@citeo\@tempcntb\m@ne\@citea\def\@citea{,}{\bf
?}\@warning
       {Citation `\@citeb' on page \thepage \space undefined}}%
    {\setbox\z@\hbox{\global\@tempcntc0\csname b@\@citeb\endcsname\relax}%
     \ifnum\@tempcntc=\z@ \@citeo\@tempcntb\m@ne
       \@citea\def\@citea{,}\hbox{\csname b@\@citeb\endcsname}%
     \else
      \advance\@tempcntb\@ne
      \ifnum\@tempcntb=\@tempcntc
      \else\advance\@tempcntb\m@ne\@citeo
      \@tempcnta\@tempcntc\@tempcntb\@tempcntc\fi\fi}}\@citeo}{#1}}
\def\@citeo{\ifnum\@tempcnta>\@tempcntb\else\@citea\def\@citea{,}%
  \ifnum\@tempcnta=\@tempcntb\the\@tempcnta\else
   {\advance\@tempcnta\@ne\ifnum\@tempcnta=\@tempcntb \else
\def\@citea{--}\fi
    \advance\@tempcnta\m@ne\the\@tempcnta\@citea\the\@tempcntb}\fi\fi}
\catcode`@=12

\begin{document}
\title{
\vskip-3cm{\baselineskip14pt
\centerline{\normalsize DESY 01-202\hfill ISSN 0418-9833}
\centerline{\normalsize hep-ph/0112259\hfill}
\centerline{\normalsize December 2001\hfill}}
\vskip1.5cm
Evidence for Colour-Octet Mechanism from CERN LEP2 $\gamma\gamma\to J/\psi+X$
Data}
\author{{\sc M. Klasen, B.A. Kniehl, L.N. Mihaila, M. Steinhauser}\\
{\normalsize II. Institut f\"ur Theoretische Physik, Universit\"at
Hamburg,}\\
{\normalsize Luruper Chaussee 149, 22761 Hamburg, Germany}}

\date{}

\maketitle

\thispagestyle{empty}

\begin{abstract}
We present theoretical predictions for the transverse-momentum distribution of
$J/\psi$ mesons promptly produced in $\gamma\gamma$ collisions within the
factorization formalism of nonrelativistic quantum chromodynamics, including
the contributions from both direct and resolved photons, and we perform a
conservative error analysis.
The fraction of $J/\psi$ mesons from decays of bottom-flavoured hadrons is
estimated to be negligibly small.
New data taken by the DELPHI Collaboration at LEP2 nicely confirm these
predictions, while they disfavour those obtained within the traditional
colour-singlet model.

\medskip

\end{abstract}

\newpage

Since its discovery in 1974, the $J/\psi$ meson has provided a useful
laboratory for quantitative tests of quantum chromodynamics (QCD) and, in
particular, of the interplay of perturbative and nonperturbative phenomena.
The factorization formalism of nonrelativistic QCD (NRQCD) \cite{bbl} provides
a rigorous theoretical framework for the description of heavy-quarkonium
production and decay.
This formalism implies a separation of short-distance coefficients, which can 
be calculated perturbatively as expansions in the strong-coupling constant
$\alpha_s$, from long-distance matrix elements (MEs), which must be extracted
from experiment.
The relative importance of the latter can be estimated by means of velocity
scaling rules, {\it i.e.}, the MEs are predicted to scale with a definite
power of the heavy-quark ($Q$) velocity $v$ in the limit $v\ll1$.
In this way, the theoretical predictions are organized as double expansions in
$\alpha_s$ and $v$.
A crucial feature of this formalism is that it takes into account the complete
structure of the $Q\overline{Q}$ Fock space, which is spanned by the states
$n={}^{2S+1}L_J^{(c)}$ with definite spin $S$, orbital angular momentum $L$,
total angular momentum $J$, and color multiplicity $c=1,8$.
In particular, this formalism predicts the existence of color-octet (CO)
processes in nature.
This means that $Q\overline{Q}$ pairs are produced at short distances in
CO states and subsequently evolve into physical, color-singlet (CS) quarkonia
by the nonperturbative emission of soft gluons.
In the limit $v\to 0$, the traditional CS model (CSM) \cite{ber,bai,hum} is
recovered.
The greatest triumph of this formalism was that it was able to correctly 
describe \cite{ebr,cho} the cross section of inclusive charmonium
hadroproduction measured in $p\overline{p}$ collisions at the Fermilab
Tevatron \cite{abe}, which had turned out to be more than one order of
magnitude in excess of the theoretical prediction based on the CSM.

In order to convincingly establish the phenomenological significance of the
CO processes, it is indispensable to identify them in other kinds of
high-energy experiments as well.
Studies of charmonium production in $ep$ photoproduction, $ep$ and $\nu N$
deep-inelastic scattering, $e^+e^-$ annihilation, $\gamma\gamma$ collisions,
and $b$-hadron decays may be found in the literature; see Ref.~\cite{bra} and
references cited therein.
Furthermore, the polarization of charmonium, which also provides a sensitive
probe of CO processes, was carefully investigated \cite{ben,bkv,bkl}.
None of these studies was able to prove or disprove the NRQCD factorization
hypothesis.

Very recently, the DELPHI Collaboration has presented preliminary data on the
inclusive cross section of $J/\psi$ photoproduction in $\gamma\gamma$
collisions ($e^+e^-\to e^+e^-J/\psi+X$) at CERN LEP2, taken as a function of
the $J/\psi$ transverse momentum ($p_T$) \cite{delphi}.
The $J/\psi$ mesons were identified through their decays to $\mu^+\mu^-$
pairs, and events where the system $X$ contains a prompt photon were
suppressed by requiring that at least four charged tracks were reconstructed.
The luminosity-weighted average $e^+e^-$ center-of-mass (c.m.) energy was
$\sqrt s=197$~GeV, the scattered positrons and electrons were antitagged, with
maximum angle $\theta_{\rm max}=32$~mrad, and the $\gamma\gamma$ c.m.\ energy
was constrained to be $W\le35$~GeV in order to reject the major part of the 
non-two-photon events.
The total cross section was found to be
$\sigma(e^+e^-\to e^+e^-J/\psi+X)=(45.3\pm18.8)$~pb.

In this Letter, we seize the opportunity to confront this data with up-to-date
theoretical predictions based on NRQCD and the CSM, in order to find out if it
is able to discriminate between the two.
In want of the full next-to-leading-order (NLO) corrections, it is
indispensable to perform a comprehensive and conservative analysis of the
theoretical uncertainties.
Doing this, we shall find that this data clearly favors the NRQCD prediction,
while the CSM one significantly falls short of it.

The photons can interact either directly with the quarks participating in the
hard-scattering process (direct photoproduction) or via their quark and gluon
content (resolved photoproduction).
Thus, the process $e^+e^-\to e^+e^-J/\psi+X$ receives contributions from the
direct, single-resolved, and double-resolved channels.
All three contributions are formally of the same order in the perturbative
expansion and must be included.
This may be understood by observing that the parton density functions (PDFs)
of the photon have a leading behavior proportional to
$\alpha\ln(M^2/\Lambda_{\rm QCD}^2)\propto\alpha/\alpha_s$, where $\alpha$ is
Sommerfeld's fine-structure constant, $M$ is the factorization scale, and
$\Lambda_{\rm QCD}$ is the asymptotic scale parameter of QCD.

In $\gamma\gamma$ collisions, $J/\psi$ mesons can be produced directly; or via
radiative or hadronic decays of heavier charmonia, such as $\chi_{cJ}$ and
$\psi^\prime$ mesons; or via weak decays of $b$ hadrons.
The respective decay branching fractions are
$B(\chi_{c0}\to J/\psi+\gamma)=(0.66\pm0.18)\%$,
$B(\chi_{c1}\to J/\psi+\gamma)=(27.3\pm1.6)\%$,
$B(\chi_{c2}\to J/\psi+\gamma)=(13.5\pm1.1)\%$,
$B(\psi^\prime\to J/\psi+X)=(55\pm5)\%$, and
$B(B\to J/\psi+X)=(1.16\pm0.10)\%$ \cite{pdg}.
The OPAL Collaboration has recently measured the total cross section of
open-bottom production in $\gamma\gamma$ collisions at LEP2, under similar
kinematic conditions as DELPHI \cite{delphi} ($\sqrt s=194$~GeV,
$\theta_{\rm max}=33$~mrad, and $10\le W\le60$~GeV), to be
$\sigma(e^+e^-\to e^+e^-b\overline{b}+X)=(14.2\pm5.9)$~pb \cite{opal}.
The cross section for $J/\psi$ mesons from $b$-hadron decays may thus be 
estimated to be $(0.33\pm0.14)$~pb, which is less than 1\% of the total
$J/\psi$ cross section measured by DELPHI and can be safely neglected.
The cross sections of the four residual indirect production channels may be
approximated by multiplying the direct-production cross sections of the
respective intermediate charmonia with their decay branching fractions to
$J/\psi$ mesons.

Invoking the Weizs\"acker-Williams approximation \cite{wei} and the
factorization theorems of the QCD parton model \cite{dew} and NRQCD
\cite{bbl}, the differential cross section of $e^+e^-\to e^+e^-H+X$, where
$H$ denotes a generic charmonium state, can be written as
\begin{eqnarray}
\lefteqn{d\sigma(e^+e^-\to e^+e^-H+X)=\int dx_+f_{\gamma/e}(x_+)\int dx_-}
\nonumber\\
&&{}\times
f_{\gamma/e}(x_-)
\sum_{a,b,d}\int dx_af_{a/\gamma}(x_a,M)\int dx_bf_{b/\gamma}(x_b,M)
\nonumber\\
&&{}\times
\sum_n\langle{\cal O}^H[n]\rangle
d\sigma(ab\to c\overline{c}[n]+d),
\label{e1}
\end{eqnarray}
where $f_{\gamma/e}(x_\pm)$ is the equivalent number of transverse photons
radiated by the initial-state positrons and electrons \cite{fri},
$f_{a/\gamma}(x_a,M)$ are the PDFs of the photon,
$\langle{\cal O}^H[n]\rangle$ are the MEs of the $H$ meson,
$d\sigma(ab\to c\overline{c}[n]+d)$ are the differential partonic cross 
sections,
the integrals are over the longitudinal-momentum fractions of the emitted
particles w.r.t.\ the emitting ones, and it is summed over
$a,b=\gamma,g,q,\overline{q}$ and $d=g,q,\overline{q}$, with $q=u,d,s$.
To leading order in $v$, we need to include the $c\overline{c}$ Fock states
$n={}^3\!S_1^{(1)},{}^1\!S_0^{(8)},{}^3\!S_1^{(8)},{}^3\!P_J^{(8)}$ if
$H=J/\psi,\psi^\prime$ and $n={}^3\!P_J^{(1)},{}^3\!S_1^{(8)}$ if
$H=\chi_{cJ}$, where $J=0,1,2$.
With the definition $f_{\gamma/\gamma}(x_\gamma,M)=\delta(1-x_\gamma)$,
Eq.~(\ref{e1}) accommodates the direct, single-resolved, and double-resolved 
channels.
The presence of parton $d$ is to ensure that $p_T$ can take finite values;
if it were absent, then $p_T$ would essentially be zero, so that only the 
lowest bin of the DELPHI data could be described.

We analytically calculated the cross sections of all contributing partonic
subprocesses and compared our results with the literature.\footnote{%
Leaving aside obvious typographical errors, we disagree
with Eq.~(7) and the subsequent equation in Ref.~\cite{jap} and
with Eq.~(4) in Ref.~\cite{ko}.}
Specifically, these include
$\gamma\gamma\to c\overline{c}[{}^3\!S_1^{(8)}]g$
\cite{ma,jap,npb};
$\gamma g\to c\overline{c}[{}^3\!S_1^{(1)}]g$ \cite{ber},
$c\overline{c}[8]g$ \cite{bkv,ko,hao};
$\gamma q\to c\overline{c}[8]q$ \cite{bkv,ko};
$gg\to c\overline{c}[{}^3\!S_1^{(1)}]g$ \cite{bai},
$c\overline{c}[{}^3\!P_J^{(1)}]g$ \cite{hum},
$c\overline{c}[8]g$ \cite{cho,bkv};
$gq\to c\overline{c}[{}^3\!P_J^{(1)}]q$ \cite{bai},
$c\overline{c}[8]q$ \cite{cho,bkv}; and
$q\overline{q}\to [{}^3\!P_J^{(1)}]g$ \cite{bai},
$c\overline{c}[8]g$ \cite{cho,bkv},
where $n=8$ collectively denotes
$n={}^1\!S_0^{(8)},{}^3\!S_1^{(8)},{}^3\!P_J^{(8)}$.
In the limit $p_T\to0$, some of these cross sections are plagued by collinear
and infrared singularities.
This happens whenever the respective $2\to1$ partonic subprocess,
$ab\to c\overline{c}[n]$, exists, namely, for
$\gamma\gamma\to c\overline{c}[{}^3\!P_{0,2}^{(1)}]$,
$\gamma g\to c\overline{c}[8^\prime]$ \cite{ko,hao,cac},
$gg\to c\overline{c}[8^\prime]$ \cite{pet}, and
$q\overline{q}\to c\overline{c}[{}^3\!S_1^{(8)}]$ \cite{pet},
where $n=8^\prime$ stands for $n={}^1\!S_0^{(8)},{}^3\!P_{0,2}^{(8)}$.
In a full NLO analysis, these singularities would be factorized at scale $M$
and absorbed into the bare PDFs and MEs so as to renormalize the latter.
In our LO analysis, we refrain from presenting predictions for the $p_T$
distribution in the lowest $p_T$ bin.
Instead, we consider the cross section arising from the $2\to1$ partonic 
subprocesses.

In our numerical analysis, we use $m_c=(1.5\pm0.1)$~GeV, $\alpha=1/137.036$,
and the lowest-order (LO) formula for $\alpha_s^{(n_f)}(\mu)$ with $n_f=3$
active quark flavors.
As for the photon PDFs, we use the LO set from Gl\"uck, Reya, and Schienbein
(GRS) \cite{grs}, which is the only available one that is implemented in the
fixed-flavor-number scheme, with $n_f=3$.
We choose the renormalization and factorization scales to be $\mu=\xi_\mu m_T$
and $M=\xi_M m_T$, respectively, where $m_T=\sqrt{(2m_c)^2+p_T^2}$ is the
transverse mass of the $H$ meson, and independently vary the scale parameters
$\xi_\mu$ and $\xi_M$ between 1/2 and 2 about the default value 1.
As for the $J/\psi$, $\chi_{cJ}$, and $\psi^\prime$ ME's, we adopt the set
determined in Ref.~\cite{bkl} by fitting the Tevatron data \cite{abe} using
the LO proton PDFs from Martin, Roberts, Stirling, and Thorne (MRST98LO)
\cite{mrst} as our default and the one referring to the LO proton PDFs from
the CTEQ Collaboration (CTEQ5L) \cite{cteq} for comparison (see Table~I in
Ref.~\cite{bkl}).
In the first (second) case, we employ $\Lambda_{\rm QCD}^{(3)}=204$~MeV
(224~MeV), which corresponds to $\Lambda_{\rm QCD}^{(4)}=174$~MeV \cite{mrst}
(192~MeV \cite{cteq}), so as to conform with the fit \cite{bkl}.
Incidentally, the GRS photon PDFs are also implemented with
$\Lambda_{\rm QCD}^{(3)}=204$~MeV \cite{grs}.
In the cases $\psi=J/\psi,\psi^\prime$, the fit results for
$\langle{\cal O}^\psi[{}^1\!S_0^{(8)}]\rangle$ and
$\langle{\cal O}^\psi[{}^3\!P_0^{(8)}]\rangle$ are 
strongly correlated, and one is only sensitive to the linear combination
\begin{equation}
M_r^\psi=\langle{\cal O}^\psi[{}^1\!S_0^{(8)}]\rangle
+\frac{r}{m_c^2}
\langle{\cal O}^\psi[{}^3\!P_0^{(8)}]\rangle,
\label{eq:mr}
\end{equation}
with an appropriate value of $r$ \cite{cho,bkv,bkl}.
Since Eq.~(\ref{e1}) is sensitive to a different linear combination of
$\langle{\cal O}^\psi[{}^1\!S_0^{(8)}]\rangle$ and
$\langle{\cal O}^\psi[{}^3\!P_0^{(8)}]\rangle$ than 
appears in Eq.~(\ref{eq:mr}), we write
$\langle{\cal O}^\psi[{}^1\!S_0^{(8)}]\rangle=\kappa
M_r^\psi$
and
$\langle{\cal O}^\psi[{}^3\!P_0^{(8)}]\rangle=(1-\kappa)
\left(m_c^2/r\right)M_r^\psi$ and vary $\kappa$ between 0 and 1 about the
default value 1/2.
The $J$-dependent MEs
$\langle{\cal O}^\psi[{}^3\!P_J^{(8)}]\rangle$,
$\langle{\cal O}^{\chi_{cJ}}[{}^3\!P_J^{(1)}]\rangle$, 
and
$\langle{\cal O}^{\chi_{cJ}}[{}^3\!S_1^{(8)}]\rangle$
satisfy the multiplicity relations collected in Eq.~(4) of Ref.~\cite{ano},
which follow to leading order in $v$ from heavy-quark spin symmetry.
In order to estimate the theoretical uncertainties in our predictions, we
vary the unphysical parameters $\xi_\mu$, $\xi_M$, and $\kappa$ as indicated
above, take into account the experimental errors on $m_c$, the decay branching
fractions, and the default MEs, and switch from our default ME set to the
CTEQ5L one, properly adjusting $\Lambda_{\rm QCD}^{(3)}$.
We then combine the individual shifts in quadrature, allowing for the upper
and lower half-errors to be different.

\begin{figure}[ht]
\begin{center}
\epsfxsize=16cm
\epsffile[0 0 567 567]{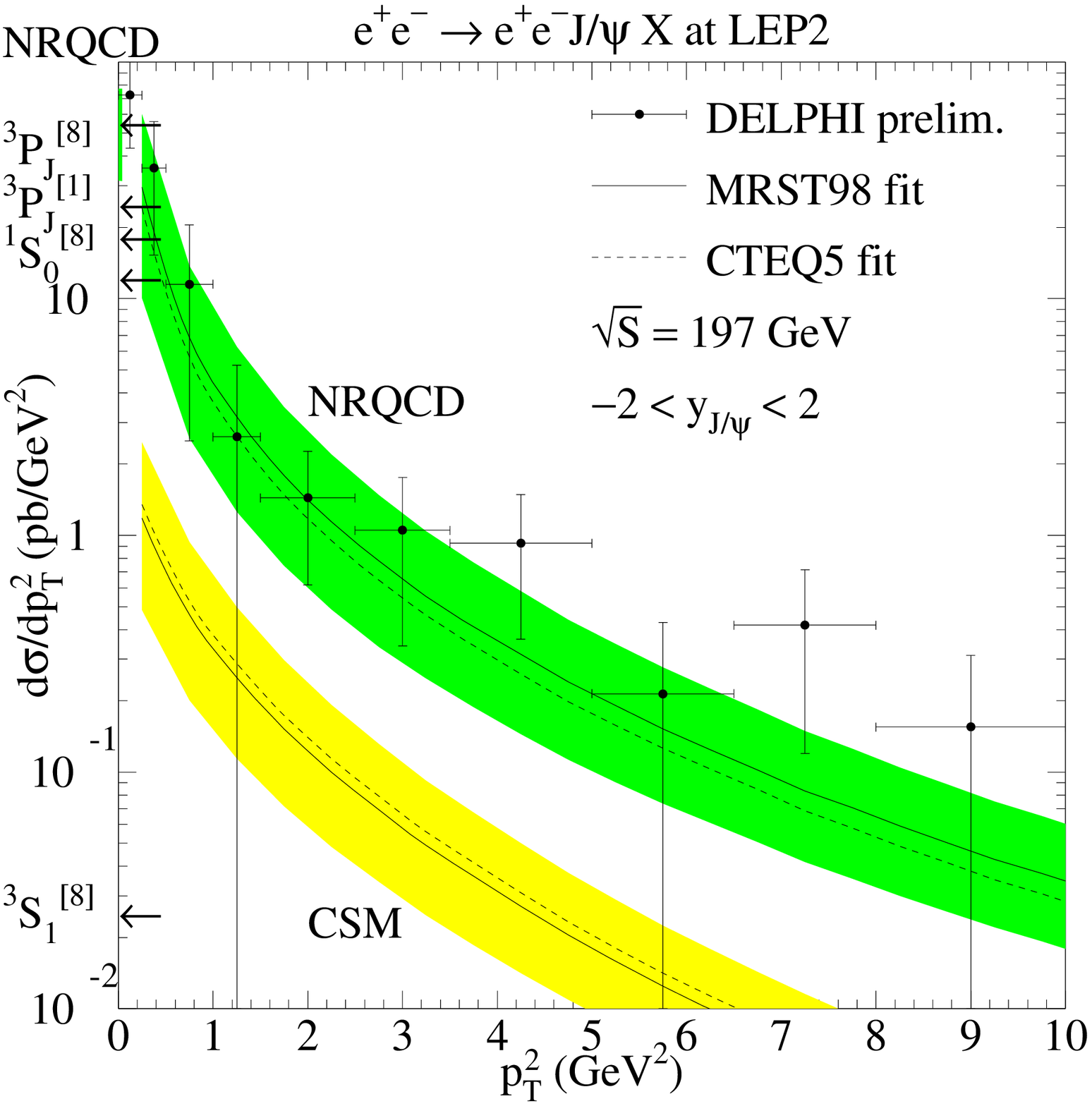}
\smallskip
\caption{The cross section $d\sigma/dp_T^2$ of $e^+e^-\to e^+e^-J/\psi+X$
measured by DELPHI \protect\cite{delphi} as a function of $p_T^2$ is compared
with the theoretical predictions of NRQCD and the CSM.
The solid and dashed lines represent the central predictions obtained with the
ME sets referring to the MRST98LO \protect\cite{mrst} (default) and CTEQ5L
\protect\cite{cteq} PDFs, respectively, while the shaded bands indicate the
theoretical uncertainties on the default predictions.
The arrows indicate the NRQCD prediction for $p_T=0$ and its 
${}^3\!P_J^{(1)}$, ${}^1\!S_0^{(8)}$, ${}^3\!S_1^{(8)}$, and ${}^3\!P_J^{(8)}$
components.}
\label{f1}
\end{center}
\end{figure}

\begin{figure}[ht]
\begin{center}
\epsfxsize=16cm
\epsffile[0 0 567 567]{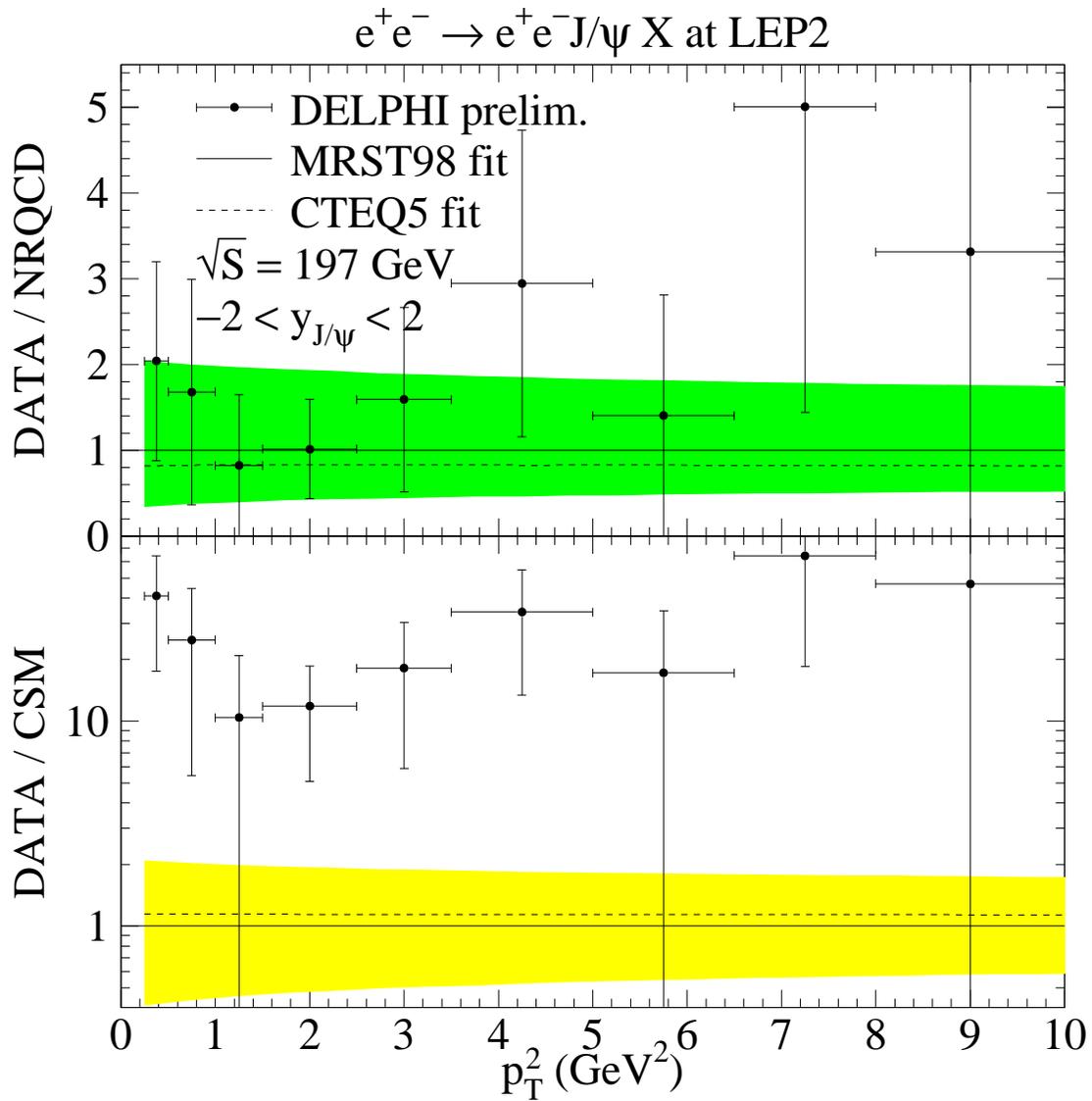}
\smallskip
\caption{Data-over-theory representation of Fig.~\ref{f1}.}
\label{f2}
\end{center}
\end{figure}

\begin{figure}[ht]
\begin{center}
\epsfxsize=16cm
\epsffile[0 0 567 567]{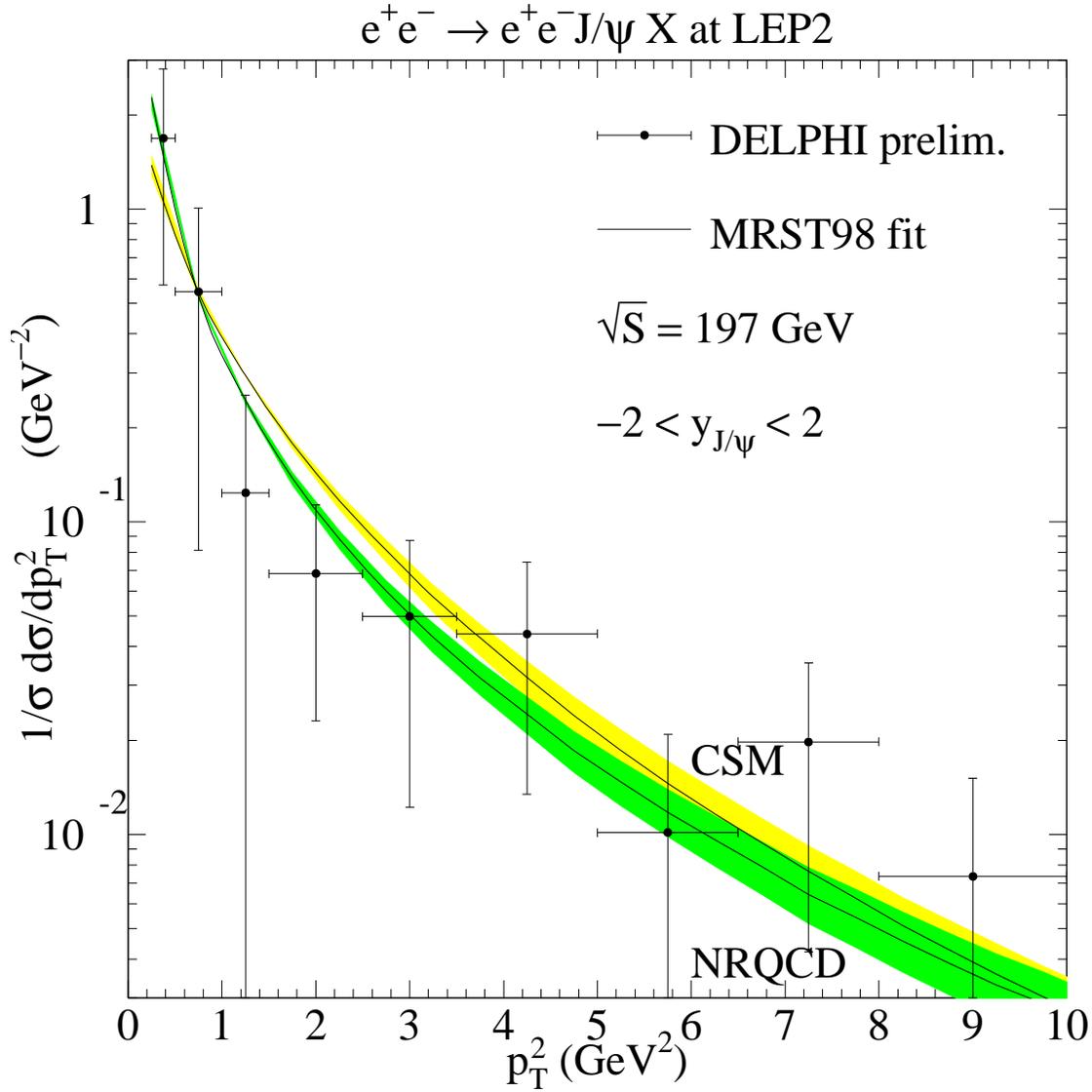}
\smallskip
\caption{The normalized cross section $(1/\sigma)d\sigma/dp_T^2$ of
$e^+e^-\to e^+e^-J/\psi+X$ measured by DELPHI \protect\cite{delphi} as a
function of $p_T^2$ is compared with the default theoretical predictions of
NRQCD and the CSM.
The shaded bands indicate the theoretical uncertainties.}
\label{f3}
\end{center}
\end{figure}

In Fig.~\ref{f1}, we confront the $p_T^2$ distribution of
$e^+e^-\to e^+e^-J/\psi+X$ measured by DELPHI \cite{delphi} with our
NRQCD and CSM predictions.
The solid lines and shaded bands represent the central results, evaluated with
our default settings, and their uncertainties, respectively.
We observe that the DELPHI data clearly favors the NRQCD prediction, while it
significantly overshoots the CSM one.
This is even more apparent from the data-over-theory representations shown in
Fig.~\ref{f2}. 
This qualitative observation can be substantiated by considering the $\chi^2$
values for the $N=9$ data points with $p_T^2\ge0.25$~GeV${}^2$.
In fact, the NRQCD central prediction yields $\chi^2/N=0.49$, which is to be
compared with 1.79 for the CSM one.
The situation is very similar for the MEs pertinent to the CTEQ5L PDFs (dashed 
lines), the corresponding results being 0.62 and 1.76, respectively.
As for the integral over the range $1\le p_T^2\le10$~GeV${}^2$, the DELPHI,
NRQCD, and CSM results read $(6.4\pm2.0)$~pb, $4.7{+1.9\atop-1.2}$~pb, and
$0.39{+0.16\atop-0.09}$~pb, respectively, {\it i.e.}, the DELPHI measurement
and the NRQCD prediction mutually agree within errors, while the CSM
prediction significantly falls short of the DELPHI result, by a factor of 16
as far as the central values are concerned.
In Fig.~\ref{f3}, we study the normalized cross section
$(1/\sigma)d\sigma/dp_T^2$, which is particularly sensitive to the shape of
the $p_T^2$ distribution and offers the advantage that the theoretical
uncertainty is greatly reduced.
We observe that both NRQCD and the CSM describe the shape of the measured
$p_T^2$ distribution well within its errors, the respective $\chi^2/N$ values
being 0.35 and 0.68, respectively.

Taking as the reference quantity the $p_T^2$ distribution integrated from 0.25
to 10~GeV${}^2$, the direct, single-resolved, and double-resolved channels 
account for 1\%, 98\%, and 1\% of the NRQCD prediction, respectively.
The situation is very similar for the CSM, except that direct photoproduction
is forbidden at LO.
In NRQCD, 91\% of the $J/\psi$ mesons are directly produced, while 9\% stem 
from the decays of $\chi_{cJ}$ and $\psi^\prime$ mesons.
In the CSM, direct production happens less frequently, in 77\% of the cases.
As explained above, the contribution from $b$-hadron decays is negligible.
Consequently, the ratio of the direct and indirect yields lends itself as a
useful discriminator between NRQCD and the CSM, and it would be desirable to
measure it in $\gamma\gamma$ collisions. 
In NRQCD, the most important error sources include the variations of
$\xi_\mu$, $m_c$, and $\kappa$, which, in average, make up 55\%, 24\%, and
11\% of the total error square
$(\delta_\uparrow\sigma)^2+(\delta_\downarrow\sigma)^2$, respectively.
In the CSM, the largest errors are related to $\xi_\mu$ (59\%), $m_c$ (30\%),
and $\xi_M$ (7\%).
It is generally believed that the magnitude of unknown higher-order
corrections may be estimated by scale variations of the known results.
For the single-resolved channel, which greatly dominates our results, the NLO
corrections in the CSM are known to be moderate at small values of $p_T$
\cite{kra}.
A similar observation was made in Ref.~\cite{npb} for the direct channel, by
studying the real radiative corrections in full NRQCD.
This indicates that the $\xi_\mu$ and $\xi_M$ variations of the LO result may
indeed provide a realistic estimate of the size of unknown higher-order
corrections.

Recent analyses \cite{sri} indicate that the overall description of the
Tevatron \cite{abe} and HERA data on inclusive charmonium production can be
improved by adopting the $k_T$ factorization approach.
It would be interesting to find out if this is also true for the DELPHI data
\cite{delphi}.

\bigskip

\noindent
{\bf Acknowledgements}

\smallskip

This work was supported in part by the Deutsche Forschungsgemeinschaft through
Grants No.\ KL~1266/1-2 and No.\ KN~365/1-1, by the Bundesministerium f\"ur
Bildung und Forschung through Grant No.\ 05~HT1GUA/4, and by Sun Microsystems
through Academic Equipment Grant No.~EDUD-7832-000332-GER.

\newpage

\newpage

\newpage

\end{document}